\documentclass[reprint,footinbib,amsmath,amssymb,%
  aps,pre,floatfix,nolongbibliography]{revtex4-2}
\usepackage{graphicx}

\newcommand{\pH}{$p\text{H}$}

\newcommand{\kB}{k_{\text{B}}}
\newcommand{\kT}{\kB T}
\newcommand{\lB}{l_{\text{B}}}
\newcommand{\lD}{\lambda_{\text{D}}}
\newcommand{\phiD}{\phi_{\text{D}}}
\newcommand{\phiW}{\phi_{\text{W}}}
\newcommand{\kabar}{{\overline\kappa}}

\newcommand{\leibnizd}{\mathrm{d}}
\newcommand{\dz}{\leibnizd z}
\newcommand{\dphi}{\leibnizd\phi}
\newcommand{\dsqphi}{\leibnizd^2\phi}

\newcommand{\Eq}[1]{Eq.~\eqref{#1}}
\newcommand{\Eqs}[1]{Eqs.~\eqref{#1}}
\newcommand{\Fig}[1]{Fig.~\ref{#1}}
\newcommand{\partFig}[2]{\Fig{#1}{#2}}

\newcommand{\naive}{{na\"\i{}ve}}

\newcommand{\latin}[1]{{\itshape #1}}
\newcommand{\cf}{\latin{cf.}}

\newcommand{\ie}{\latin{i.$\,$e.}}

\newcommand{\inextremis}{\latin{in extremis}}

\newcommand{\german}[1]{{\itshape #1}}

\newcommand{\gedankenexperiment}{\german{Gedankenexperiment}}

\begin{document}

\title{Partial osmotic pressures of ions in electrolyte solutions}

\author{Patrick B. Warren}

\email{patrick.warren@stfc.ac.uk}

\affiliation{The Hartree Centre, STFC Daresbury Laboratory, Warrington, WA4 4AD, United Kingdom}

\date{\today}

\begin{abstract}
  The concept of the partial osmotic pressure of ions in an electrolyte solution is critically examined.  In principle these can be defined by introducing a solvent-permeable wall and measuring the force per unit area which can certainly be attributed to individual ions.  Here I demonstrate that although the total wall force balances the bulk osmotic pressure as required by mechanical equilibrium, the individual partial osmotic pressures are extra-thermodynamic quantities dependent on the electrical structure at the wall, and as such they resemble attempts to define individual ion activity coefficients.  The limiting case where the wall is a barrier to only one species of ion is also considered, and with ions on both sides the classic Gibbs-Donnan membrane equilibrium is recovered thus providing a unifying treatment.  The analysis can be extended to illustrate how the electrical state of the bulk is affected by the nature of the walls and the sample handling history, thus supporting the `Gibbs-Guggenheim uncertainty principle' (the notion that the electrical state is unmeasurable and usually accidentally determined).  Since this uncertainty is conferred also onto individual ion activities, it has implications for the current (2002) IUPAC definition of \pH.
\end{abstract}

\maketitle

\section{Introduction}
The concept of individual ion activities, tentatively introduced by Lewis and Randall in 1923 \cite{LR23}, continues to provoke fierce debates to the present day.  On the one hand Guggenheim, following earlier work by Gibbs, came to the conclusion in 1929 that individual ion activities must be regarded as ill-defined quantities since they depend on the unknown electrical state of the system under consideration \cite{Gug29}.  On the other hand the current (2002) IUPAC definition of \pH\ \cite{BRC+02, *Baucke2002} as the negative base-10 logarithm of the hydrogen ion activity would seem to place undue emphasis on what many workers would regard as an `extra-thermodynamic' quantity.

The essential problem is expressed by what one might term the \emph{Gibbs-Guggenheim uncertainty principle} \cite{Pethica2007, *Hall1978}.  This is the notion that the electrical state of a solution is not only unmeasurable but also `usually accidentally determined' to boot \cite{Gug67}.  To account for this Guggenheim introduced the concept of the electrochemical potential, in which the unknown electrical state is represented by what is effectively the mean electrostatic potential of the bulk system.  Building on this, it follows that only the mean activities of neutral combinations are thermodynamically well-defined, since the mean electrostatic potential cancels out \cite{Gug67, HW72}.  Most workers adhere to this paradigm, and assorted proposals to define and measure individual ion activities never seem to survive deeper scrutiny \cite{WVV05, *Mal06, AWVV07, *Mal10b, *WVAV10, ZDW+08, *Mal10a, RAWV09, *Mal09, DZZ+10, *Mal11b, *DZZ+11, FM11, *Mal11a, *Fer11, Zar11, *VWV12a, *Zar12a, *VWV12b, *Zar12b, Fra12a, *Zar12c, *Fra12b}.

\begin{figure}[b]
  \centering
  \includegraphics[clip=true,width=0.8\columnwidth]{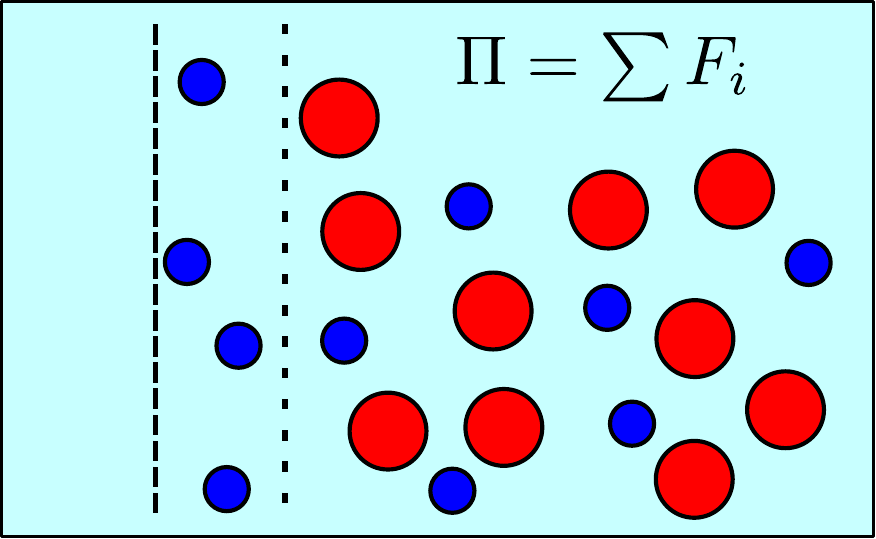}
  \caption{In an electrolyte the forces $F_i$ per unit area that can be ascribed to individual ion species depend on the nature of the wall (\ie\ they are extra-thermodynamic quantities), although summed they must equal the bulk osmotic pressure $\Pi$.\label{fig:cartoon}}
\end{figure}

In modern parlance, one might say that the mean electrostatic potential is determined by what happens at the walls, and as such is not a bulk thermodynamic property.  This is as true in computer simulations as it is in real systems, and precludes the \naive\ use of individual ion activities for parametrisation and model coarse-graining.  This spoils an otherwise attractive proposition since for $N$ ionic species it would render an $O(N^2)$ problem (fitting data for all possible neutral combinations) into an $O(N)$ problem (fitting only individual species data).

Perhaps instead one can use the partial osmotic pressures for ions? These are easily measured in simulations by computing the mean force per unit area exerted by individual ion species at a solvent-permeable wall introduced for this purpose (\Fig{fig:cartoon}) \cite{NR19}.  It appears they might provide the sought-after $O(N)$ advantage that eludes the individual ion activities, but without the accompanying thermodynamic `baggage' \cite{WVV11a, *WVV11b, *Zar13, *Fer13, *Roc15, *VB15, *FF16, *WVV16, *Kak20}.  But here I argue that in an electrolyte solution these partial osmotic pressures are likewise extra-thermodynamic quantities.  Although the wall force can certainly be decomposed into contributions from individual species, and the total must match the bulk osmotic pressure, the individual contributions are dependent on the electrical structure at the wall and the adopted wall model.

To demonstrate this I shall introduce a number of `toy models' in which the role played by the electrical structure at the wall is made explicit. Armed with these it is then possible to build toy models of containers which demonstrate explicitly the origin of the Gibbs-Guggenheim uncertainty principle.  For simplicity and analytic convenience, I shall use hard walls and Poisson-Boltzmann (PB) theory to calculate the ion density profiles and the corresponding wall forces.  These models can be adapted to the case where the wall is a barrier to only one ion species, or is ion-selective.  With this, the classic Gibbs-Donnan membrane equilibrium can be recovered, thus providing a unifying treatment.  The use of toy models is not a limitation as such.  Rather, if problems arise in these cases for what are clearly identifiable reasons, it is obvious that they must also arise in more realistic or more complex models, including in computer simulations.  

\section{Toy models of walls}
In approaching these problems I shall consider the example of a 1:1 electrolyte treated within the PB approximation \cite{Isr11, SW02, RvR10, Xin11, VBB+12, MV14, *MV16}. For the most part I will work with reduced units $q=\kT=1$, where $q$ is the fundamental unit of charge, $\kB$ is Boltzmann's constant, and $T$ is the temperature.  I first consider the case where the wall is modelled as a generic pair of repulsive potentials and demonstrate that whilst the sum of the forces is always equal to the bulk osmotic pressure, as required by mechanical equilibrium, the individual contributions depend on the details of the potentials.  Thus although the partial osmotic pressures of individual ion species can be defined, they are extra-thermodynamic quantities in the sense that they depend on the wall potential.

As a warm-up exercise let me outline an even simpler model discussed by Marbach and Bocquet \cite{MB19, LA12}.  Suppose one has an ideal gas of particles at a density $\rho(z)$, subject to a repulsive potential $U(z)$, with $U(z)\to0$ and $\rho\to\rho_s$ as $z\to\infty$, and $\rho\to0$ as $z\to-\infty$, so the gas is bounded from the left hand side.  The grand potential and density thereof are, respectively,
\begin{equation}
\Omega=\int_{-\infty}^{\infty}\!\omega\,\dz\,,\quad
\omega=\rho\Bigl(\ln\frac{\rho}{\rho_s}-1\Bigr)+\rho U\,.
\label{eq:Oint}
\end{equation}
As $z\to\infty$, one has $\omega\to-\rho_s$ thus identifying $\Pi=\rho_s$ as the bulk osmotic pressure from the Gibbs-Duhem relation.  From the variational principle $\delta\Omega/\delta\rho(z)=0$ one finds that the particle density is Boltzmann-distributed, $\rho(z)=\rho_s e^{-U}$.  The force on the wall is then given by
\begin{equation}
  \begin{split}
  F&=\int_{-\infty}^\infty\Bigl(-\frac{\partial U}{\partial z}\Bigr)\,\rho\,\dz
  =  \rho_s\int_{-\infty}^\infty
  \frac{\partial (e^{-U})}{\partial z}\,\dz\\[3pt]
  &{}=\rho_s \bigl[e^{-U}\bigr]_{-\infty}^\infty=\rho_s\,.
  \end{split}
\end{equation}
Thus the force on the wall is equal to the osmotic pressure, as should be the case from the point of view of mechanical equilibrium.

I now extend this to a 1:1 electrolyte within the PB approximation, which treats the ions as an electrostatically interacting but otherwise ideal gas of positive and negative point charges at densities $\rho_\pm(z)$.  I shall suppose that there is a \emph{pair} of repulsive potentials $U_\pm(z)$ which act \emph{separately} on each species of ion.  The grand potential is again given by the integral in \Eq{eq:Oint} but now the grand potential density is \cite{SW02}
\begin{equation}
  \omega=\sum_{i=\pm}\rho_i\Bigl(\ln\frac{\rho_i}{\rho_s}-1\Bigr)
  +\frac{E^2}{8\pi\lB}+\rho_+U_++\rho_-U_-
\end{equation}
where $E=-\partial\phi/\partial z$ is the electric field in reduced units, and $\phi$ is the dimensionless electrostatic potential (\ie\ in units of $\kT/q$ as mentioned).  This latter quantity satisfies the Poisson equation,
\begin{equation}
  \frac{\partial^2\phi}{\partial z^2}+4\pi\lB\rho_z=0\,,\label{eq:poiss}
\end{equation}
where $\rho_z=\rho_+-\rho_-$ is the net charge density and $\lB$ is the Bjerrum length (restoring units, $\lB=q^2\!/\epsilon\kT$ where $\epsilon$ is the dielectric permittivity).

I shall suppose that $U\pm\to0$ and $\rho_\pm\to\rho_s$ as $z\to\infty$, and that $\rho_\pm\to0$ as $z\to-\infty$ (this latter constraint will be relaxed in the final case study below).  As above, one identifies from this that the bulk ($z\to\infty$) osmotic pressure is $\Pi=2\rho_s$ as befits the presence of two species of ions at equal densities.  The variational principle applied to this problem yields again a Boltzmann distribution for the two ion species, $\rho_\pm(z)=\rho_se^{\mp\phi}e^{-U_\pm}$.

I now consider the two wall forces separately, thus for example
\begin{equation}
  \begin{split}
    F_+&=\int_{-\infty}^\infty\Bigl(-\frac{\partial U_+}{\partial z}\Bigr)\,
    \rho_+\,\dz
  =  \rho_s\int_{-\infty}^\infty\!\!
  e^{-\phi}\,\frac{\partial(e^{-U_+})}{\partial z}\,\dz\\[3pt]
  &{}=\bigl[\rho_se^{-\phi}e^{-U_+}\bigr]_{-\infty}^\infty
  +\rho_s\int_{-\infty}^\infty\!\!
  e^{-\phi}e^{-U_+}\,\frac{\partial\phi}{\partial z}\,\dz
  \end{split}
\end{equation}
(integrating by parts).
Evaluating this, and making a similar calculation for $F_-$, results in
\begin{equation}
  F_\pm=\rho_s\pm\int_{-\infty}^\infty\!\!\rho_\pm\,
  \frac{\partial\phi}{\partial z}\,\dz\,.
  \label{eq:fsep}
\end{equation}
The total force is therefore
\begin{equation}
  F = F_++F_-=2\rho_s+\int_{-\infty}^\infty\!\!\rho_z
    \frac{\partial\phi}{\partial z}\,\dz\,.
\end{equation}
The latter integral here vanishes by virtue of \Eq{eq:poiss},
\begin{equation}
  \begin{split}
    \int_{-\infty}^\infty\!\!\rho_z
    \frac{\partial\phi}{\partial z}\,\dz
    &=-\frac{1}{4\pi\lB}\int_{-\infty}^\infty
    \frac{\partial^2\phi}{\partial z^2}\,
    \frac{\partial\phi}{\partial z}\,\dz\\[3pt]
    &\qquad{}=-\frac{1}{8\pi\lB}\Bigl[\Bigl(
      \frac{\partial\phi}{\partial z}\Bigr)^{\!2}\,
      \Bigr]_{-\infty}^\infty=0
  \end{split}
\end{equation}
(the electric field $\partial\phi/\partial z\to0$ as $z\to\pm\infty$).
Thus one sees that $F_++F_-=2\rho_s$ and as in the previous case the total force balances the total osmotic pressure.

The second term in \Eq{eq:fsep} clearly shows that the electrical structure at the wall transfers force from one type of ion to the other.  This makes it abundantly clear that the individual wall forces \emph{must} depend on the (arbitrary) choice of wall potentials, and it is only the total force that matches the bulk osmotic pressure.

\begin{figure}
  \centering
  \includegraphics[clip=true,width=\columnwidth]{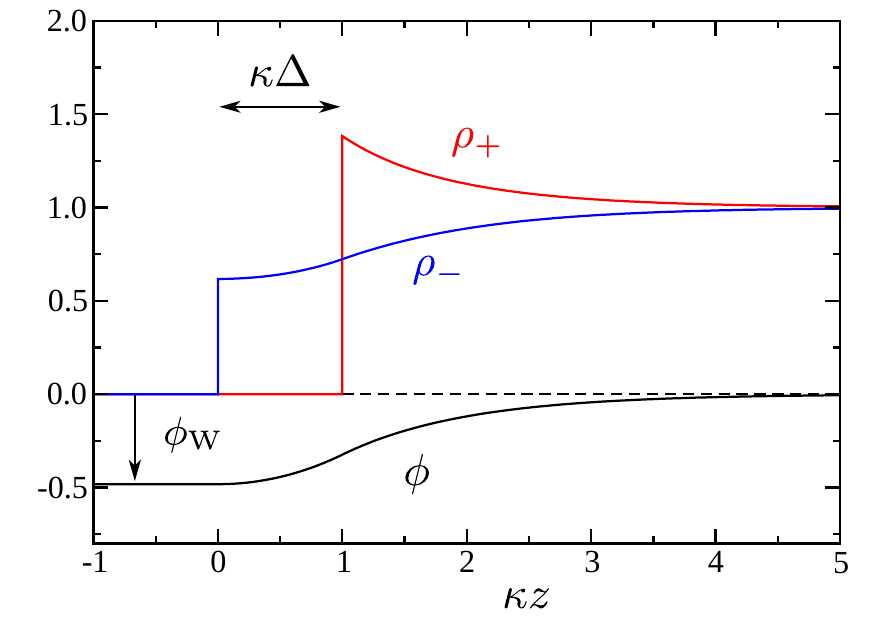}
  \caption{An electrolyte solution with a split hard wall potential.  Shown are the ion density profiles $\rho_{\pm}(z)$ and the electrostatic potential $\phi(z)$.  The problem is solved for $\rho_s=\kappa\Delta=1$ for which $\phiW\approx-0.483\approx-12.4\,\mathrm{mV}$ (room temperature).\label{fig:split_profile}}
\end{figure}

To reinforce the above analysis, with a suitable choice of $U_\pm(z)$ one can calculate the ion density profiles analytically and solve explicitly for the force that each ion exerts on its respective potential barrier.  Here, I consider a split pair of hard repulsive potential barriers (\Fig{fig:cartoon}).  This may reflect for instance a model in which the ions have different diameters so that their centres of mass are excluded at different distances.  In this model therefore $U_\pm=\infty$ for $z<z_\pm$ and $U_\pm=0$ for $z>z_\pm$, where for concreteness and without loss of generality I shall suppose $z_-<z_+$. Thus there are two hard barriers to the ions, separated from each other by a distance $\Delta=z_+-z_-$.  The ion densities obey $\rho_\pm(z)=0$ for $z<z_\pm$ and $\rho_\pm=\rho_s e^{\mp\phi}$ for $z>z_\pm$.  Injecting the corresponding charge density $\rho_z=\rho_+-\rho_-$ into the Poisson equation, \Eq{eq:poiss}, yields (PB equation),
\begin{equation}
  \frac{\dsqphi}{\dz^2}=\Bigg\{\begin{array}{lll}
  0 && z<z_-\,,\\
  4\pi\lB\rho_s e^\phi && z_-<z<z_+\,,\\
  8\pi\lB\rho_s\sinh\phi && z>z_+\,.
  \end{array}
\end{equation}
It will be convenient to introduce $\kappa^2=8\pi\lB\rho_s$ so that $\lD=\kappa^{-1}$ corresponds to the Debye length defined in the bulk ($z\to\infty$).  A first integration of the above gives
\begin{equation}
  \frac{\dphi}{\dz}=\Bigg\{\begin{array}{lll}
  0 && z<z_-\,,\\
  \kappa({e^\phi-e^{\phi_-}})^{1/2} && z_-<z<z_+\,,\\
  -2\kappa\sinh(\phi/2) && z>z_+\,,
  \end{array}
\end{equation}
where $\phi_\pm=\phi(z_\pm$) and continuity across $z=z_-$ has been imposed.  For continuity across $z=z_+$ one should have
\begin{equation}
({e^{\phi_+}-e^{\phi_-}})^{1/2}=-2\sinh(\phi_+/2)\label{eq:cont1}
\end{equation}
which can be reduced to $e^{-\phi_+}+e^{\phi_-}=2$.

\begin{figure}
  \centering
  \includegraphics[clip=true,width=\columnwidth]{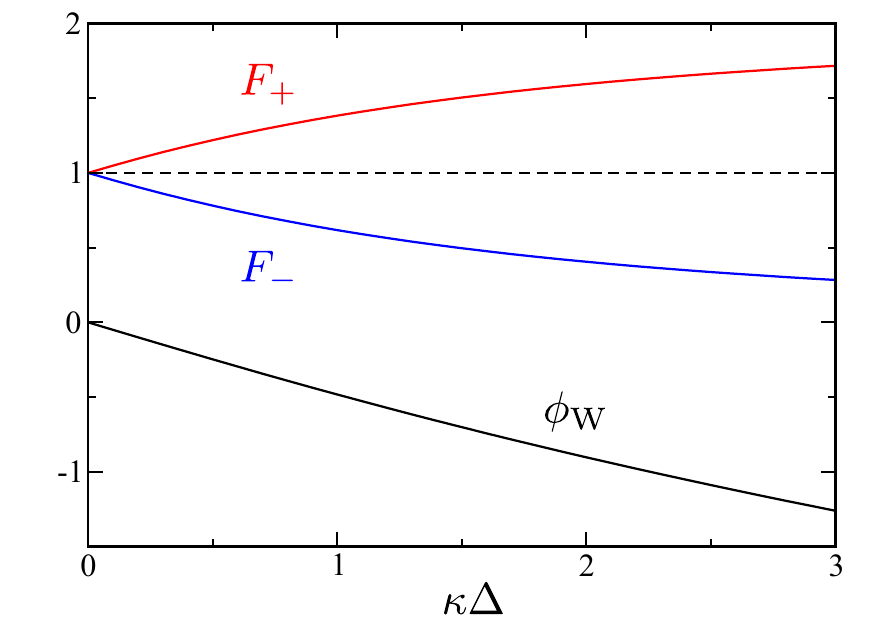}
  \caption{The wall forces $F_\pm$ from individual ions, and the electrostatic wall potential $\phiW$, as a function of the dimensionless separation $\kappa\Delta$ in a split potential model with $\rho_s=1$.\label{fig:split_wallforce}}
\end{figure}

A second integration now yields
\begin{equation}
  \phi=\Bigg\{\begin{array}{lll}
  \phi_- \>(\,\equiv \phiW) && z<z_-\,,\\[3pt]
  \phi_--\ln\cos^2\bigl[e^{\phi_-/2}\kappa(z-z_-)/2\bigr]
  && z_-<z<z_+\,,\\[3pt]
  4\tanh^{-1}\bigl[\tanh(\phi_+/4)e^{-\kappa(z-z_+)}\bigr]
  && z>z_+\,,
  \end{array}\label{eq:EDL}
\end{equation}
where I have introduced the electrostatic `wall potential', $\phiW$, being the difference between the electrostatic potential in the exterior region ($z<z_-$) and that in  the bulk electrolyte solution ($z\to\infty$).

To anyone familiar with the literature, \Eq{eq:EDL} will be recognisable as a stitching together of two classic textbook solutions to the PB equation \cite{Isr11, RvR10, VBB+12, MV14, MV16}.  In writing the above, continuity across $z_-$ has again been assumed. Imposing continuity across $z_+$ requires $\exp({\phi_--\phi_+})=\cos^2(e^{\phi_-/2}\kappa\Delta/2)$.  With the aid of $e^{-\phi_+}+e^{\phi_-}=2$ from \Eq{eq:cont1}, this can be reduced to a transcendental equation for $x\equiv e^{\phi_-/2}$ in terms of $\kappa\Delta$,
\begin{equation}
  x^4-2x^2+\cos^2(x\kappa\Delta/2)=0\,.
\end{equation}
This provides a complete solution to the problem.  An example is shown in \Fig{fig:split_profile} for $\kappa\Delta=1$ for which $x\approx0.786$.

It remains to provide an expression for the wall forces.  Inserting the Boltzmann-distributed ion density profiles into \Eq{eq:fsep} finds that $F_\pm=\rho_s e^{\mp\phi_\pm}$, in other words the forces are given by the respective contact densities at the hard walls.  By virtue of the first of the above continuity conditions, the sum $F_++F_-=2\rho_s$ as claimed earlier. \Fig{fig:split_wallforce} shows the two forces as a function as $\kappa\Delta$.  As the potential barriers move further apart, more and more of the total force is carried by the positive ions, which build up in front of the leading potential barrier as indicated in \Fig{fig:split_profile}.  At the same time the electrostatic wall potential (also shown in \Fig{fig:split_wallforce}) becomes increasingly negative.  Note that the total charge in the electrical double layer (EDL) remains zero by virtue of Gauss' principle since there is no electric field for $z<z_-$ nor in the bulk electrolyte. This can also be discovered by integrating the Poisson equation, $4\pi\lB \int_{z_-}^\infty\! \rho_z\,\dz =-[{\dphi}/{\dz}]_{z_-}^\infty=0$.

\begin{figure}
  \centering
  \includegraphics[clip=true,width=\columnwidth]{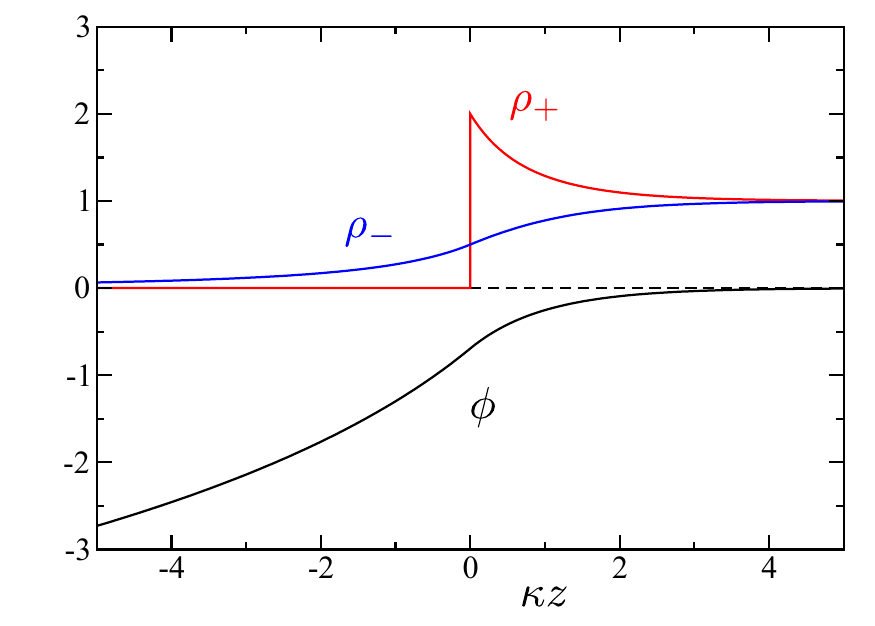}
  \caption{Ion density profiles and electrostatic potential for a hard wall that blocks only the positive ions, equivalent to \Fig{fig:split_profile} in the limit $\kappa\Delta\to\infty$. Plots are for $\rho_s=1$.\label{fig:single_profile}}
\end{figure}

As a limiting case, one can let the barrier separation tend to infinity, in which case there is a single wall acting on only one species of ion. This case is of interest because it illustrates \inextremis\ how the total osmotic pressure in the bulk is transmitted to the wall by just one of the ionic species.  Let us select the positive ions and set the hard repulsive barrier at $z=0$.  The PB equation for this case is then
\begin{equation}
  \frac{\dsqphi}{\dz^2}=\bigg\{\begin{array}{lll}
  4\pi\lB\rho_s e^\phi && z<0\,,\\
  8\pi\lB\rho_s\sinh\phi && z>0\,.
  \end{array}
\end{equation}
The first integration gives
\begin{equation}
  \frac{\dphi}{\dz}=\bigg\{\begin{array}{lll}
  \kappa e^{\phi/2} && z<0\,,\\
  -2\kappa\sinh(\phi/2) && z>0\,,
  \end{array}
\end{equation}
assuming that $\dphi/\dz\to0$ as $z\to-\infty$.  Continuity across $z=0$ requires $\kappa e^{\phi_0/2}=-2\kappa\sinh(\phi_0/2)$ where $\phi_0=\phi(0)$.  This can be solved to obtain $\phi_0=-\ln2$.  Integrating once more gives the full solution
\begin{equation}
  \phi=\bigg\{\begin{array}{lll}
  -2\ln(\sqrt{2}-\kappa z/2) && z<0\,,\\[3pt]
  4\tanh^{-1}(2\sqrt{2}-3)e^{-\kappa z}) && z>0\,.
  \end{array}
\end{equation}
A plot of this solution is shown in \Fig{fig:single_profile}. Unlike the previous case or the next case, $\phi$ diverges logarithmically as $z\to-\infty$ although $\dphi/\dz$ vanishes asymptotically as $-(\kappa z)^{-1}$.  The force on the wall is due to the confinement of the positive ions alone, $F_+=\rho_s e^{-\phi_0}$. Using the above result for $\phi_0$, one sees that $F_+=2\rho_s$, which as claimed fully accounts for the bulk osmotic pressure.

\begin{figure}
  \centering
  \includegraphics[clip=true,width=\columnwidth]{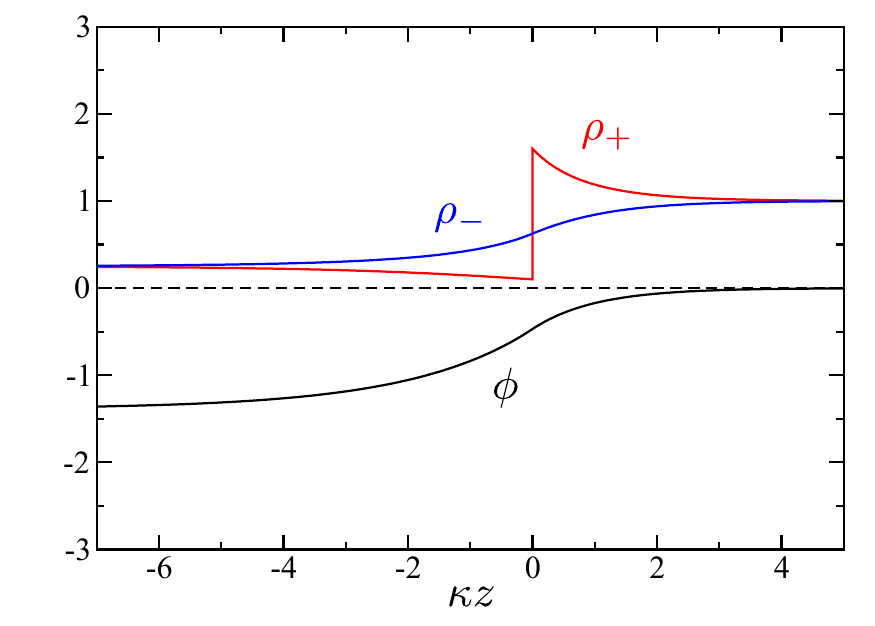}
  \caption{Ion density profiles and electrostatic potential for a step potential of magnitude $U_0=4\ln2$ that acts only on the positive ions.  Plots are for $\rho_s=1$.  Restoring units, the Donnan potential $\phiD = -U_0/2 \approx -35.6\, \mathrm{mV}$ (room temperature).\label{fig:donnan_profile}}
\end{figure}

To complete the hat-trick as it were, I now consider what happens in the above problem if the wall becomes a potential step of a \emph{finite} height.  In this case both species of ions can cross, establishing a Gibbs-Donnan membrane equilibrium.  For this problem, for the positive ions there is a potential barrier of the form $U_+=U_0>0$ for $z<0$ and $U_+=0$ for $z>0$; whereas for the negative ions $U_-=0$ everywhere. The ion densities then satisfy
\begin{equation}
  \rho_+=\bigg\{\begin{array}{ll}
  \rho_s e^{-\phi} e^{-U_0} & z<0\,,\\
  \rho_s e^{-\phi} & z>0\,,
  \end{array}\qquad
  \rho_-=\rho_s e^{\phi}\quad\forall\, z\,.
\end{equation}
Making use of the expectation that the ion densities should become equal to one another as $z\to-\infty$, one concludes that in this limit $\phi\to\phiD=-U_0/2$ (the `Donnan potential').  The PB equation for this problem can then be written as
\begin{equation}
  \frac{\dsqphi}{\dz^2}=\bigg\{\begin{array}{lll}
  \kabar^2\sinh(\phi-\phiD) && z<0\,,\\
  \kappa^2\sinh\phi && z>0\,.
  \end{array}
\end{equation}
where $\kabar^2=\kappa^2e^{\phiD}$ (\cf\ \cite{GvRK01}).  The first integral is
\begin{equation}
  \frac{\dphi}{\dz}=\bigg\{\begin{array}{lll}
  2\kabar\sinh((\phi-\phiD)/2) && z<0\,,\\[3pt]
  -2\kappa\sinh(\phi/2) && z>0\,.
  \end{array}
\end{equation}
Continuity across $z=0$ then determines the potential at the step as $\phi_0=\ln[(1+e^{\phiD})/2]$ from which the full solution can be constructed as a pair of back-to-back EDLs as in the last of \Eqs{eq:EDL},
\begin{equation}
  \phi=\Bigg\{\begin{array}{lll}
  \phiD+4\tanh^{-1}\bigl[\tanh((\phi_0-\phiD)/4)
    e^{\kabar z}\bigr] && z<0\,,\\[3pt]
  4\tanh^{-1}\bigl[\tanh(\phi_0/4)e^{-\kappa z}\bigr] && z>0\,.
  \end{array}
\end{equation}
An example is shown in \Fig{fig:donnan_profile}, where the potential step $U_0=4\ln2\approx2.77$ is chosen so that the asymptotic ion densities on the left hand side are one quarter of the asymptotic ion densities on the right hand side.

Finally, by an extension of the analysis for the above two cases, the force on the step is given by the difference in the contact values for the positive ions, namely $F_+=\rho_+(0^+)-\rho_+(0^-)=\rho_s e^{-\phi_0}(1-e^{2\phiD})$.  Making use of the continuity condition to eliminate $\phi_0$ reduces this to $F_+=2\rho_s(1-e^{\phiD})$. The osmotic pressure on the right hand side ($z\to\infty$) is $2\rho_s$ as before, and the osmotic pressure on the left hand side ($z\to-\infty$) is $2\rho_s e^{\phiD}$ reflecting the reduction in the ion densities on that side.  Therefore $F_+$ is equal to the difference in these osmotic pressures, as one would expect.

This limiting case reproduces in a physical model the classic Gibbs-Donnan membrane equilibrium, and is similar to a calculation reported earlier for the electrical structure at a `jellium' half space \cite{SW02}.  Note that the EDL is more compressed on the right hand side, where the asymptotic ion densities are larger; again \cf\ \cite{GvRK01}.

\begin{figure}
  \centering
  \includegraphics[clip=true,width=0.8\columnwidth]{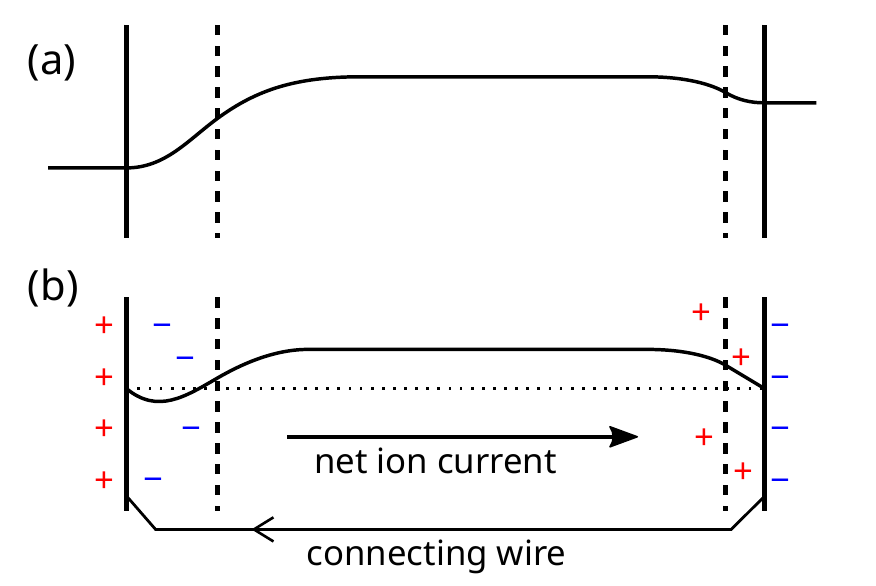}
  \caption{(a) An isolated system with different walls at either end exhibits a difference in the exterior electrostatic potentials. (b) If the walls are short-circuited a transfer of charge takes place until the exterior potential difference vanishes.  When equilibrium is re-established, the diffuse charge in the EDLs balances the wall charges, and for this to happen there must also have been a net ion current in the bulk.\label{fig:charging}}
\end{figure}

\section{Toy models of containers}
The various models discussed above, in particular the split hard wall model, show that the electrical structure at the interface depends on details at the wall.  Thus, the electrostatic wall potential also depends on these details.  This raises the interesting question about what happens if an electrolyte solution is contained in a vessel where the walls are not uniform.  In equilibrium, the mean electrostatic potential in the bulk of the electrolyte should be constant, and since $\phiW$ varies from place to place, so does the external electrostatic potential.  This implies the existence of an \emph{exterior} electric field, similar to the stray external fields that arise from facet-dependent work functions in a metal \cite{FBB02}.  This is illustrated in \Fig{fig:charging}a where the electrolyte is bound by different walls on the left and right hand sides.  Note that since the exterior electric field develops over a macroscopic distance of order the container size, it is normally utterly negligible on the length scale of the EDLs.

What happens if the walls are short-circuited?  In this case the system should behave exactly like an electrochemical cell in the sense that an electric current flows through the connecting `wire' until charges build up at the walls to compensate for the bare wall potentials (\Fig{fig:charging}b).  When equilibrium is re-established, the diffuse charge in the EDL balances the wall charge by the same argument made earlier (Gauss' principle mandates that there can be no net charge if there are no electric fields in the exterior region or in the bulk electrolyte).  Since the system starts with uncharged EDLs, a net ion current must also have flowed in the other direction through the electrolyte solution.  By analogy to other EDL charging problems \cite{BTA04}, the time scale for this charging process should then be of the order $\lD L/D$ where $\lD$ is the Debye length, $L$ is the vessel size, and $D$ is the diffusion coefficient of the ions. Inserting numbers suggests that this all takes place quite quickly, for example with $\lD \sim 10\, \text{nm}$, $L \sim 1\, \text{cm}$ and $D \sim 10^{-9}\, \text{m}^2\, \text{s}^{-1}$, the time scale is of the order 0.1\,s.

\begin{figure}
  \centering
  \includegraphics[clip=true,width=0.8\columnwidth]{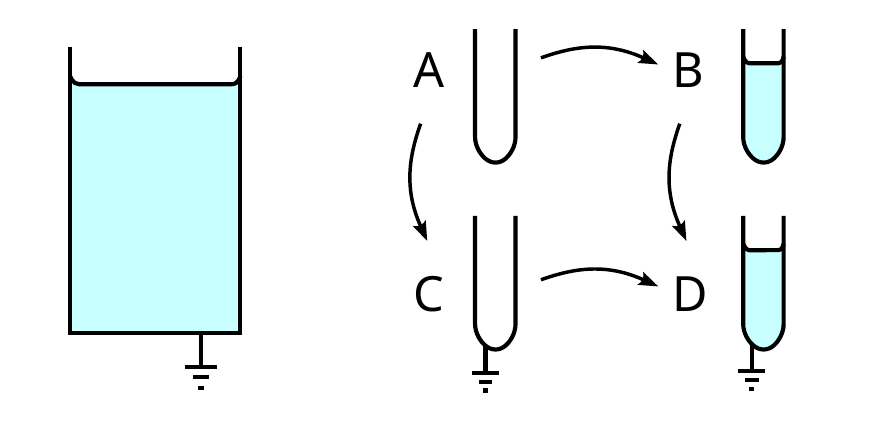}
  \caption{If a sample tube (A) is filled with an electrolyte solution from an earthed container (left), the final electrical state depends on whether the sample tube is earthed first, before filling (A$\to$C$\to$D), or filled first and then earthed (A$\to$B$\to$D).\label{fig:electroscope}}
\end{figure}

In the resulting new equilibrium, the walls carry surface charges (assuming the walls are `blocking' in the sense that no electrochemical reactions take place).  But the split wall model above assumes that the wall is uncharged ($\dphi/\dz=0$ at $z=z_-$).  This implies that the models should be extended to take into account a sheet of wall charges at $z=z_-$.  For the wall models discussed above such calculations can be performed but are rather tedious and unenlightening, and as before one ends up with a combination of textbook solutions to the PB equation.  The key point is that $\phiW$ now depends not only on the details of the wall but also on its state of charge.  For a further consideration we can also imagine that the walls may have different areas.  In this case the amount of charge that needs to be redistributed to equalise the electrostatic wall potentials depends on the relative areas of the walls, and so therefore does the final resulting $\phiW$.

The above arguments establish that the electrical state of the bulk (interior) of the electrolyte solution (captured by $\phiW$) depends on the nature of the walls of the containing vessel, but still more arbitrariness can arise.  Consider the process of filling a sample tube with electrolyte solution from an earthed container (\Fig{fig:electroscope}).  Then it matters whether the tube is earthed first before filling, or filled first and then earthed.  The process resembles the classic demonstration of charging a gold-leaf electroscope by induction \cite{Ass10}, where of course the exact sequence of operations is crucial to obtain the desired result.

The reason why the final state can be different can be traced to charging argument in \Fig{fig:charging}.  Imagine the left hand side represents the container, and the right hand side represents the sample tube.  If the tube is not initially earthed, and supposing for simplicity there is negligible charge transfer in the loading step, then the situation resembles \partFig{fig:charging}{a}.  In this case there is no net charge in the sample tube and the interior will have a mean electrostatic potential determined by the walls as indicated above.  On the other hand, if the sample tube is earthed whilst being filled, the situation more closely resembles \partFig{fig:charging}{b}.  In this case the sample tube will acquire a net charge with a current to earth being balanced by an ion current in the electrolyte solution during loading.  In the final state the interior will have a \emph{different} mean electrostatic potential, being determined by a combination of the walls and the amount of charge transferred.  Thus from this  \gedankenexperiment\ one concludes that the final electrical state of the electrolyte solution in the sample tube is sensitive not only to the walls but also to the handling history, and by extension to the history of the container, and so on.  Likewise these considerations indicate the absolute electrical state is unmeasurable, since it seems impossible to construct a protocol in which $\phiW$ would not be affected somewhere by uncontrolled wall effects and sample preparation history.

\section{Discussion}
Within the PB approximation the individual ion osmotic pressures are simply $\rho_i\kT$ where the $\rho_i$ are the individual ion densities.  However with a more complicated model, the corresponding assignment is not obvious.  Clearly though, by mechanical force balance the total force per unit area on a solvent-permeable wall must equal the bulk osmotic pressure.  Since one can always break the wall force down into the contributions from individual species, one can always write $\Pi=\sum F_i$.  It is tempting therefore to identify these individual contributions with the partial osmotic pressures of the ions.  The main purpose of the present analysis is to argue that such a decomposition is ambiguous because it depends on the electrical structure at the wall.  The $F_i$ are extra-thermodynamic quantities and in this sense they resemble attempts to define individual ion activities.

For making coarse-grained electrolyte models, since the results depend on the nature and electrical structure at the walls, it is clear that one should take care when mapping between atomistic and coarse-grained levels of description.  In principle the ambiguities can be resolved by measuring and compensating for the detailed electrical structure at the wall, but this seems to be a far from trivial task.  It might be thought that one could eliminate the problem by a judicious choice of wall model: after all, was it not a bit silly to use a different potential for the ions in the above toy models?  However, using the same potential for each ion does not guarantee the absence of wall effects if the ions have asymmetric interactions with each other, or with the solvent. These asymmetries will propagate to the ion density profiles at the wall, so that an electrical structure will inevitably develop in a similar way to the above toy models.  Such asymmetric ion-ion and ion-solvent interactions seem inevitable if a model is to capture specific ion effects such as represented by the Hofmeister series \cite{CW85}.

A secondary purpose of this work is to draw attention to the fact that the electrical state of the bulk electrolyte solution represented by the mean electrostatic potential is demonstrably an extra-thermodynamic variable.  With some simple thought experiments, it is possible to show that it not only depends on the nature of the walls, but also on the handling history of the sample.  This supports the Gibbs-Guggenheim uncertainty principle that the electrical state is unmeasurable and usually accidentally determined.

In the context of the current IUPAC definition of \pH\ \cite{MV06, deL10, Kak15, deL14}, this presents a challenge.  The unknown electrical state corresponds to an uncertainty in the mean electrostatic potential of order $\kT/q\approx 25\,\mathrm{mV}$ (at room temperature) which translates to an uncertainty $\Delta (p\mathrm{H})\approx1/\ln10\approx 0.43$ units.  This is markedly larger than the \emph{precision} with which \pH\ is defined and can be measured (typically $\pm0.01$ to $\pm0.02$ units).  Conversely, specifying the hydrogen ion activity to the indicated precision would amount to controlling the mean electrostatic potential to better than $0.2\,\mathrm{mV}$, one-tenth of the uncertainty deriving from the Gibbs-Guggenheim principle.  This problem has of course not gone unnoticed but the alternate oft-proposed approach of defining \pH\ in terms of the hydrogen ion \emph{concentration} just seems to introduce its own set of difficulties \cite{MV06}.  An actual \pH\ measurement is reduced to practice by means of a series of thermodynamically well-defined operations \cite{BRC+02, *Baucke2002, deL14}, so in a sense these difficulties ought to be purely conceptual.  Perhaps the resolution then is to introduce a distinction between the `true' single ion activity proscribed by Guggenheim, and an `apparent' single ion activity that is measurable and reflects the experimental protocols and electrochemistry underpinning the current IUPAC definition of \pH.

\begin{acknowledgments}
This work was supported by the STFC CLASP programme (grant number ST-S00646X-1).  I thank Rosalind Allen for a critical reading of an early draft of the manuscript.
\end{acknowledgments}

%\bibliography{gdliterature}
%apsrev4-2.bst 2019-01-14 (MD) hand-edited version of apsrev4-1.bst
%Control: key (0)
%Control: author (8) initials jnrlst
%Control: editor formatted (1) identically to author
%Control: production of article title (-1) disabled
%Control: page (0) single
%Control: year (1) truncated
%Control: production of eprint (0) enabled
%

\end{document}